\documentclass[a4paper,twocolumn]{article}

\usepackage{graphicx}

\title{First Order Reversal Curve Analysis on Superconductors}
\author{E. R. Di Lascio \\ \scriptsize{erlascio@protonmail.com}}
\date{\today}

\begin{document}
\twocolumn[
\begin{@twocolumnfalse}
\maketitle
\abstract

The magnetometric technique of First Order Reversal Curve (FORC) analysis, applicable to hysteretic systems, is introduced to the study of superconducting samples. Some typical superconducting structures in FORC diagram are identified, and the reversible and irreversible components are isolated, allowing the identification of typical magnetic features for superconductors and therefore extending the usefulness of the technique to more complex systems like hybrid S/N, S/F, and spin valves by improving the characterization of interactions among the components. 

\ \ \\ \ \\ \ 
\end{@twocolumnfalse}]
\section{Introduction}

First Order Reversal Curve (FORC) analysis is a magnetometric technique that is very useful in studying hysteretic systems. It has arised as a means to understand hysteron modelling \cite{Mayergoyz} and became increasingly independent from it through subsequent developments \cite{Pike, Pike2, Pike3, Martinez}, being applied to a large number of systems, mainly ferromagnetic. 

Though it has been available for more than 20 years, its interpretation is still evolving as it gains popularity. One of the main issues is that its origins, being related to hysteron modelling, make the physical interpretation of its characteristics more abstract. One has to think of the switching field for a hysteron instead of a physical element. As more people look into it, some aspects are clarified and its usefulness becomes more evident \cite{Gilbert} as well as some limitations \cite{Ruta}. 

One good way to appreciate the potential of FORC analysis is to deviate from hysteron modelling and to map the structures present in a diagram \cite{Gilbert}. These can lead to physical interpretation, like the reversible and irreversible aspects of the magnetization occurring in the same material, despite the apparent inability to isolate particular elements of material. The reversible and irreversible magnetization components are results of physical processes nonetheless. FORC analysis often allows the separation of these, thus potentially leading to important insights of processes ocurring in the material and serving as a guide to theoretical interpretation and hypothesis testing. The mapping of the magnetic characteristics present in a FORC diagram can lead to the identification of magnetic fingerprints, which can be very useful when dealing with more complex hybrid systems.

Most of the research has focused on ferromagnetic systems. But this also has potential to change, since lots of systems now lie in the interface of ferromagnetism and other materials. Acknowledging the usefulness of FORC analysis in ferromagnetic systems, the next question is, what about other materials and hybrid systems? 

Recent advances in the field of spintronics \cite{Linder}, where the spin degree of freedom of the electron is manipulated, continuously show the importance of the interaction of superconducting and ferromagnetic materials. Particularly interesting is the relation of superconducting spin currents in triplet superconductors and ferromagnetic resonace experiments \cite{Jeon_2018}. The dynamics of spin conduction, and especially the role of the superconducting vortex dynamics (annihilation, motion, entry and exit) in modifying the purposed spin current is not clear. To tackle it, one has to better understand the reversible and irreversible components of magnetization of the superconductor, therefore comprehending the hysteretic response of a superconductor and its relation with the underlying vortex dynamics (for instance \cite{Geim_1997}) is fundamental to advance the understanding of possible spin currents in superconductors. 

Another important aspect of the vortex dynamics in relation with spin currents is the flux-flow spin Hall effect \cite{Vargunin_2019}, where, unlike experiments with ferromagnetic resonance that point to a spin triplet, a spin singlet pairing is at the root of spin transport in type II superconductors. In this case the spin is transported by the magnetic vortices. Despite understanding the role of vortices in this effect, it is not clear the role of the hysteretic response of the superconductor, since it was studied close to $H _{c2}$. Nor is clear the role of a pinning array, as it changes the motion and structure of the vortex array, studied at the flux-flow regime.

Since the vortex dynamics and its control are related to the magnetic response of the superconductor, and this response is critical in recent developments in spintronics, its important to obtain one from the other and various techniques are available to numerically obtain the magnetic response from more fundamental quantities, such as the Ginzburg-Landau model \cite{Gorkov, Hu, Schmid}, Keldysh-Usadel equations \cite{Schmid2, Belzig}, or the classical vortex dynamical model \cite{Huebener}. On the other hand, coming from the magnetic measurements and trying to interpret the results in terms of vortex dynamics and possible distinct regimes is a less easy inverse problem \cite{Inverse}. 

A more detailed magnetometric study of superconductors and hybrid systems will advance with the understanding of the detailed hysteretic responses, especially if it demonstrates the existence of particular characteristic features and their relation to the underlying vortex dynamics. A detailed picture, meaning not just the stationary vortex lattice motion under a fixed magnetic field, but also vortex entry, exit and anihilation, as well as non-statinary configurations of the vortex lattice, can be obtained with a tool for practical magnetometic analysis for superconducting materials, pure or in heterostructures. This tool will be particularly useful considering the increasing complexity of the devices with hybrid materials. This paper introduces the FORC analysis in superconductors as such a tool.

The paper brings only the briefest summary of FORC analysis, since quality material are abundant elsewhere (e.g., \cite{Gilbert, Pike}). From there and a brief relation to hysteron modelling, type II superconducting samples are simulated using the Time Dependent Ginzburg Landau equations (\cite{Gorkov, Hu, Schmid}, for a review see e.g. \cite{lascio1}), with emphasis on capturing the magnetic reversal curves. The FORC diagrams for superconductors are presented, leading to the identification of typical structures that may serve as a magnetic fingerprint for superconductors, what can be particularly useful in more complex heterostructures. The samples include vortex pinning arrays, which are capable of modifying the underlying vortex dynamics. Currents are applied, to assess the effect on the magnetization of reaching a critical current. As the samples are analysed, separable structures have their hysteresis curves reconstructed and reversible and irreversible components are separated, demonstrating the usefulness of this technique to hybrid systems, where some superconducting characteristics may be superposed to characteristics of other materials, hysteretic or not.

\section{Simulations}

The results presented in this article are numeric, derived from the simulation of type II superconducting samples along the lines of \cite{lascio1}. 

The simulated samples are assumed two-dimensional, since many interesting results can be obtained with them, with good agreement with experimental results for thin films \cite{rede}. The results can be applied to a three-dimensional sample whenever border effects in the third dimension ($z$) are negligible. Since the $z$ dependence of vortices is much smaller than the $x, y$ in conventional superconductors, the two-dimensional model is justified when the demagnetization factor is negligible \cite{demagnetization}. 

The simulation proceed via TDGL (the details can be found in \cite{lascio1}). The equations representing the order parameter and the link variables that result in the magnetic field are given by:

\begin{equation}
\Delta _{j} (t + \delta t) = \Delta _{j}(t) + \mathcal{F} _{\Delta} ^{j} (t) \delta t,
\label{tdgl_delta}
\end{equation}
\begin{equation}
U _{x} ^{jk} (t + \delta t) = U _{x} ^{jk} (t) \exp(\mathcal{F} _{U _{x}} ^{jk} (t) \delta t),
\label{tdgl_ux}
\end{equation}
\begin{equation}
U _{y} ^{jm} (t + \delta t) = U _{y} ^{jm} (t) \exp(\mathcal{F} _{U _{y}} ^{jm} (t) \delta t),
\label{tdgl_y}
\end{equation}
where:
\begin{eqnarray}
\mathcal{F} _{\Delta} ^{j} (t) \equiv \frac{1}{12}\Bigg[\frac{U _{x} ^{kj} (t) \Delta _{k} (t) - 2 \Delta _{j} (t) + U _{x} ^{ij} (t) \Delta _{i} (t)} {a _{x} ^{2}} + \nonumber \\ +\frac{ U_{y} ^{mj} (t) \Delta _{m} (t) - 2 \Delta _{j} (t) + U _{y} ^{gj} (t) \Delta _{g} (t)}{a _{y} ^{2}} - \nonumber \\ - (1-T)(|\Delta _{j} (t)| ^{2} - \nu _{j})\Delta _{j} (t)\Bigg] + \tilde{f} _{j} (t), \nonumber
\end{eqnarray}
\begin{eqnarray}
\mathcal{F} _{U _{x}} ^{jk} (t) \equiv -i(1-T)\mbox{Im}\{\Delta _{k} ^{*} (t) U _{x} ^{jk} (t) \Delta _{j} (t)\} - \nonumber \\ - \Big(\frac{\kappa ^{2}}{a _{y} ^{2}}\Big)[U _{y} ^{kn} (t) U _{x} ^{nm} (t) U _{y} ^{mj} (t) U _{x} ^{jk} (t) \cdot \nonumber \\ \cdot U _{y} ^{kh} (t) U _{x} ^{hg} (t) U _{y} ^{gj} (t) U _{x} ^{jk} (t) -1],\nonumber
\end{eqnarray}
\begin{eqnarray}
\mathcal{F} _{U _{y}} ^{jm} (t) \equiv -i(1-T)\mbox{Im}\{\Delta _{m} ^{*} (t) U _{y} ^{jm} (t) \Delta _{j} (t)\} - \nonumber \\ - \Big(\frac{\kappa ^{2}}{a _{x} ^{2}}\Big)[U _{x} ^{ml} (t) U _{y} ^{li} (t) U _{x} ^{ij} (t) U _{y} ^{jm} (t) \cdot \nonumber \\ \cdot U _{x} ^{mn} (t) U _{y} ^{nk} (t) U _{x} ^{kj} (t) U _{y} ^{jm} (t) -1]. \nonumber
\end{eqnarray}

These equations are the result of the integration of the TDGL equations, discretized in space, using the Euler method. The number of vertices used is $N _{x} = N _{y} = 95$. This represents a sample of size $\sim 0.5 \mu \mbox{m} \times 0.5 \mu \mbox{m}$ for a superconductor with $\xi (0) = 100$ \AA.

Throughout the article, the units used are normalized (\cite{Hu}). The temperature is measured in units of $T _{c}$, and the lower and upper critical fields result, at $T = 0.5$, $H _{c1} = 0.04$ and $H _{c2} = 0.5$. The value of the Ginzburg-Landau parameter used is $\kappa = 2$.

\section{FORC analysis}

Magnetic hysteresis are closed loops in the relation between the magnetization of a given material and the applied magnetic field. From it one can usually obtain mangnetometric relevant quantities such as bias, remanence and coercive field.

\begin{figure}[!htb]
\centering
\includegraphics[width=0.5\textwidth]{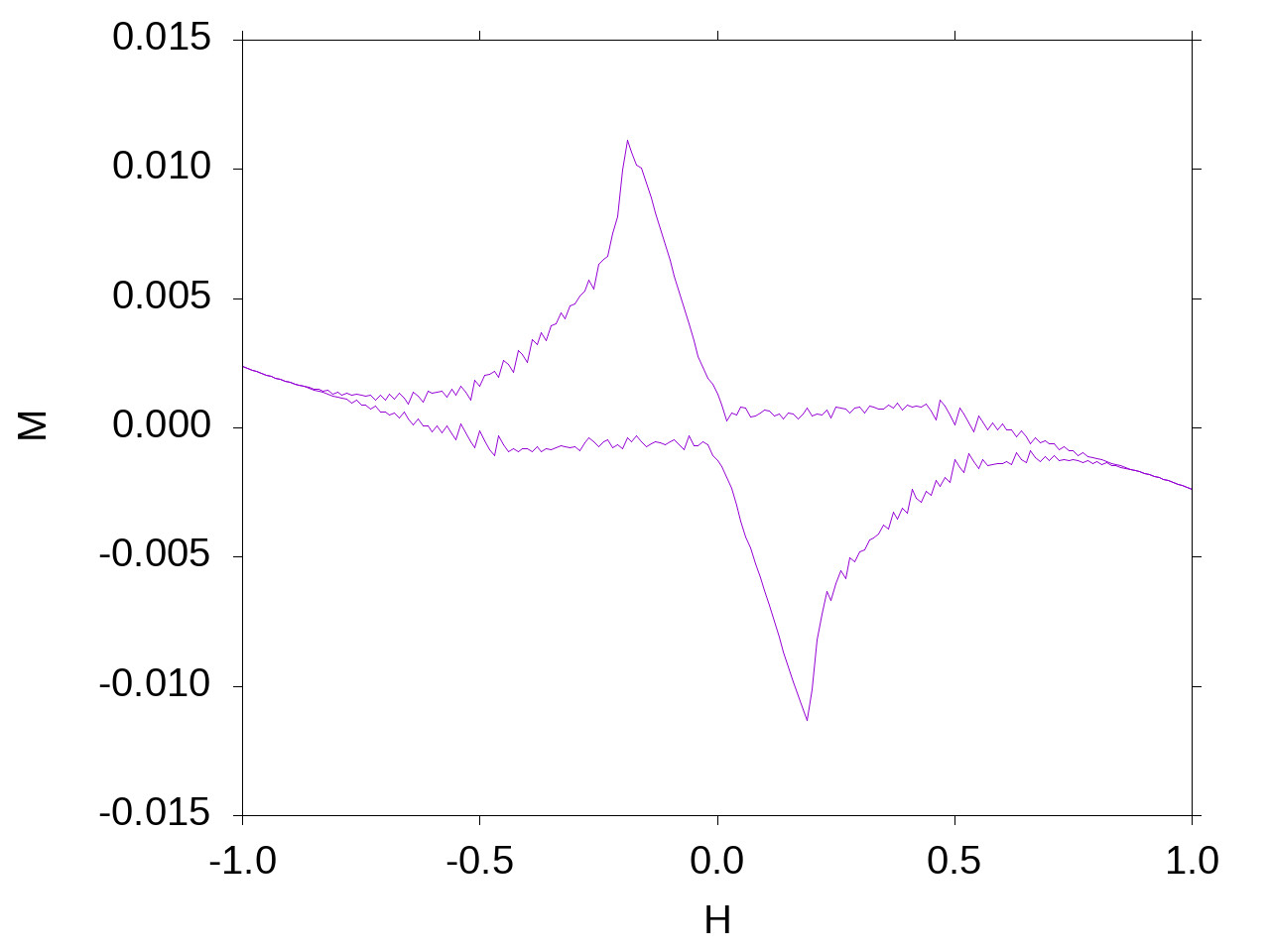}
\caption{Hysteresis of sample without defects.}
\label{hysteresis_sd_0}
\end{figure}

Superconductors present this phenomenon as well (see for instance \cite{Geim_1997}). It is related to with the irreversibility of vortex dynamics, and in the case of a sample without pinning of vortices, the main contribution comes from the surface (Bean-Livingston \cite{Bean}) barrier, though even for the sample without pinning array, the remanence is not zero. Considering the simplest sample described in the previous section, representing a square slab of type II superconducting material without defects (pinning centers), the magnetic hysteresis obtained with TDGL is exemplified by figure \ref{hysteresis_sd_0}.

Comparing with a typical (e.g. ferromagnetic) loop, the hysteresis seems reversed. With higher applied fields, the magnetization gets more negative, which is a consequence of the Meissner effect. The bias is zero, the remanence is $m _{r} = 1.3 \cdot 10^{-3}$ and the coercive field is $h _{c} = -0.59$, though it is worth emphasizing that the magnetization is positive for very negative values of the field and as the field increases, it reaches zero and continues to fall. 

As will be seen shortly, FORC analysis depends, among other things, on the definition of a saturation value. It is worth noting that the hysteresis vanishes (the magnetization values for increasing and decreasing fields are the same) not far from $H _{c2}$, but there is no saturation (in the sense of a horizontal asymptote). After that, the magnetization follows a path compatible with a diamagnetic material, but never truly saturates. Hence, when a saturation value is needed for FORC analysis, it will be taken as an arbitraty point in the magnetization curve where, regardless of increasing or decreasing the field, the magnetization follows the same path. That is, locally there is no hysteresis, or the magnetization is locally reversible for the point taken as saturation, for the purposes of FORC analysis.

While the hysteresis curve contains a lot of information, First Order Reversal Curve (FORC) analysis is a technique that allows probing the whole area of the hysteresis, not only its border, thus giving more information about processes going on the material as the field is varied. 

FORC analysis allows the identification and characterization of magnetic phases and their separation, along with characterization of interactions among components, thus allowing relevant insights into the microscopic details based on macroscopic magnetometric measurements.  

Consider a point in the upper branch of the major hysteresis loop obtained by varying the field from positive saturation (say, $H = 1$) to a reversal field (say, $H _{r} = 0$) and measurements of magnetization as the field is increased back to saturation. The curve so obtained is called a FORC (or a FORC branch). This curve probes, for the elements that had their magnetization switched while  going from saturation to reversal field, the switching field back to the initial state. 

In a ferromagnet, for instance, that would mean the elements that had their magnetization reversed from the initial state of saturation, and the FORC probing the switching field for each of those elements, as the field returns to the saturation value (for instance, see \cite{Gilbert}). In type II superconductors that can mean vortex entry, annihilation, exit. 

Hence, the derivative:
$$
\frac{\partial}{\partial H} M(H, H_{r})
$$
gives the contribution to the magnetization of each up-switching event whilst moving in a given FORC (identified by the reversal field $H _{r}$).

Now, considering a small variation in the reversal field, ${H '} _{r} = H _{r} - \delta H$, ($\delta H > 0$), repeating the procedure and comparing both curves, the difference gives the new up-switching events. That is, decreasing $H _{r}$ has increased the number of down-switching events accumulated, which will then be switched while the field is increased back to saturation.

So, while the $H$ variation probes up-switching events, variations in $H _{r}$ probes down-switching events.

Combining both sources of variation, one obtains the FORC distribution:

\begin{equation}
\rho(H, H _{r}) = -\frac{1}{2} \frac{\partial}{\partial H _{r}} \left( \frac{\partial }{\partial H } M (H, H _{r}) \right).
\label{forc_distribution}
\end{equation}

This function arises in the Preisach-Krasnoselskii model, as demonstrated by Mayergoyz \cite{Mayergoyz}. The factor of 2 in the denominator appeared afterwards, and is related to a normalization condition (see \cite{Pike2}). Besides, the saturation magnetization is often seen in the denominator, in the definition \ref{forc_distribution}. Here it will be omitted, but the only difference is the normalization value obtained upon integration. 

The distribution \ref{forc_distribution}, when integrated over $H$ and $H _{r}$, gives the saturation magnetization ($M _{S}$) if the reversible component is included, as shown in \cite{Pike2}. For this to happen, the FORCs must be extended. The procedure used for that is keeping the magnetization constant, for $H < H _{r}$, for a given FORC. That is:

\begin{equation}
M(H, H _{r}) = \left\{
\begin{array}{ll}
M(H,H _{r}) \mbox{\hspace{0.3cm} if \hspace{0.3cm}} H \geq H _{r} \\
M(H _{r}, H_{r}) \mbox{\hspace{0.2cm} if \hspace{0.2cm}} H < H _{r}
\end{array}
\right.
\label{forc_ext}
\end{equation} 

The relation between variation of $H, H_{r}$ and up and down switching events allows obtaining the ascending and descending branches of the major hysteresis loop, as shown in \cite{Gilbert}. Using the relation of $H _{r}$ with down-switching events, one can recover the descending branch of the major hysteresis loop upon integration:

\begin{equation}
M(H _{d}) = M _{S} - 2 \int _{-\infty} ^{\infty} \int _{H _{d}} ^{\infty} \rho (H, H _{r}) d H _{r} d H
\label{descending_branch}
\end{equation}

Considering the relation of $H$ and up-switching events, one obtains the ascending branch:
\begin{equation}
M(H _{u}) = - M _{S} + 2 \int _{-\infty} ^{H _{u}} \int _{-\infty} ^{\infty} \rho (H, H _{r}) d H _{r} d H
\label{ascending_branch}
\end{equation}

As shown in \cite{Gilbert}, the recovery of hysteresis branches upon integration is even more useful in combination with the FORC distribution capability of allowing identification of magnetic behavior in multiphase systems. This will be used further to identify characteristic structures (magnetometric "fingerprints") of superconductors and their relation to magnetic behavior, thus allowing the application of the technique to more complex hybrid systems. 

As is the case for ferromagnets, the FORC analysis has no spatial resolution, so it can't be associated to a particular physical component or, for instance, a particular vortex entering the sample. It models mathematical hysteretic elements (hysterons \cite{Mayergoyz}), and the challenge of the FORC technique is to relate this hysteron modelling to physical processes happening in the sample.

The algorithm to obtaining FORC data is straightforward. The sample is initially brought to positive saturation $H_{S}$. The field is then reduced to a reversal field $H_{r}$, and is subsequently increased back to saturation by incremental $\Delta H$ steps, while the magnetization is measured at each step. Once reaching saturation, the process is repeated with $H_{r} \rightarrow H_{r} - \Delta H_{r}$, until $H_{r} = - H_{S}$. Each curve so obtained is refered to as FORC, or alternatively as a FORC branch, and is associated with each value of $H_{r}$. In this article, the usual choice of $\Delta H = \Delta H _{r}$ will be employed.

In the simulations presented in this article, the field interval to be swept goes from $H = 1$ to $H = -1$. Given the computing power available, 100 FORCs were generated for each sample, which gives the field step (or the maximum field resolution chosen) as $\Delta H = 0.02$. 

\begin{figure}[!htb]
\centering
\includegraphics[width=0.5\textwidth]{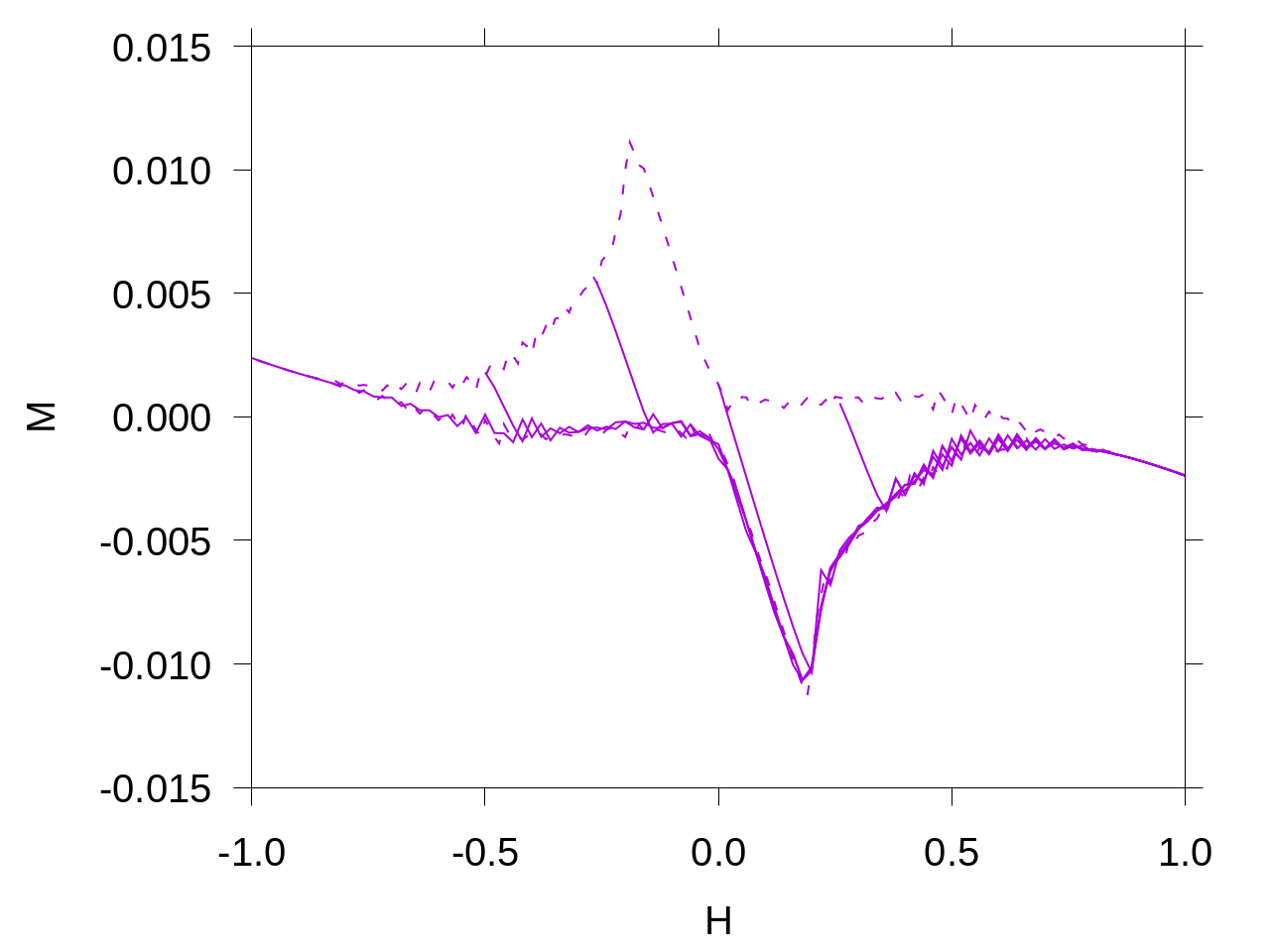}
\caption{Five FORCs for the sample without defects. The dashed line is the major hysteresis loop, shown here as a guide to the eye.}
\label{fiveforcs_sd_0}
\end{figure}

Considering the same sample from figure \ref{hysteresis_sd_0}, obtaining the FORC would be exemplified by the figure \ref{fiveforcs_sd_0}. The reversal curves start at different positions in the major hysteresis loop (shown here as a dashed line to guide the eye) and vary until reaching $H = 1$. 

The figure is a simplification to illustrate the process, since to obtain a FORC distribution many curves are necessary (as mentioned before, in this article, 100 curves were simulated for each sample). The magnetization was extended according to \ref{forc_ext} to capture the reversible component. From there, the FORC distribution is obtained according to eq. \ref{forc_distribution}.

In order to numerically calculate the distribution (eq. \ref{forc_distribution}), one usually fits a second degree polynomial to the data in a square array around the reference point \cite{Pike}. 

To generate the dataset for the polynomial estimation, one selects an array of points in the discrete $(H, H _{r})$ space, with the reference point at its center. The number n is chosen to be odd, with $n = 2\cdot SF + 1$, where $SF$ is called a smoothing factor, and $n \times n$ points are used, so that each point demands a regression with $n \times n - 6$ degrees of freedom. 
 
The polynomial to be fit takes the form:

\begin{eqnarray}
M(H, H _{r}) = \alpha + \beta _{1} \cdot H + \beta _{2} \cdot H _{r} + \nonumber \\ + \beta _{3} \cdot H ^{2} + \beta _{4} \cdot H _{r} ^{2} + \beta _{5} \cdot H \cdot H _{r}
\label{polynomial_forc}
\end{eqnarray}

This represents the polynomial expansion up to the first non-zero element after the derivatives in eq. \ref{forc_distribution} are calculated. The distribution in the point $H, H _{r}$ is then taken to be $-0.5 \cdot \beta _{5}$. An alternative calculation is shown in the Appendix.

\section{Results}

The samples simulated in this study were slabs of superconductor. One of the samples was simulated without defects, but arrays of defects were introduced to assess the effect of pinning centers in the sample, giving also a hint as to the behavior changes expected with hybrid materials. The defects arrays were square, triangular, center and border concentrated (see \cite{lascio1}). Simulations with a non-zero applied current value were also performed.

\begin{figure}[!htb]
\centering
\includegraphics[width=0.5\textwidth]{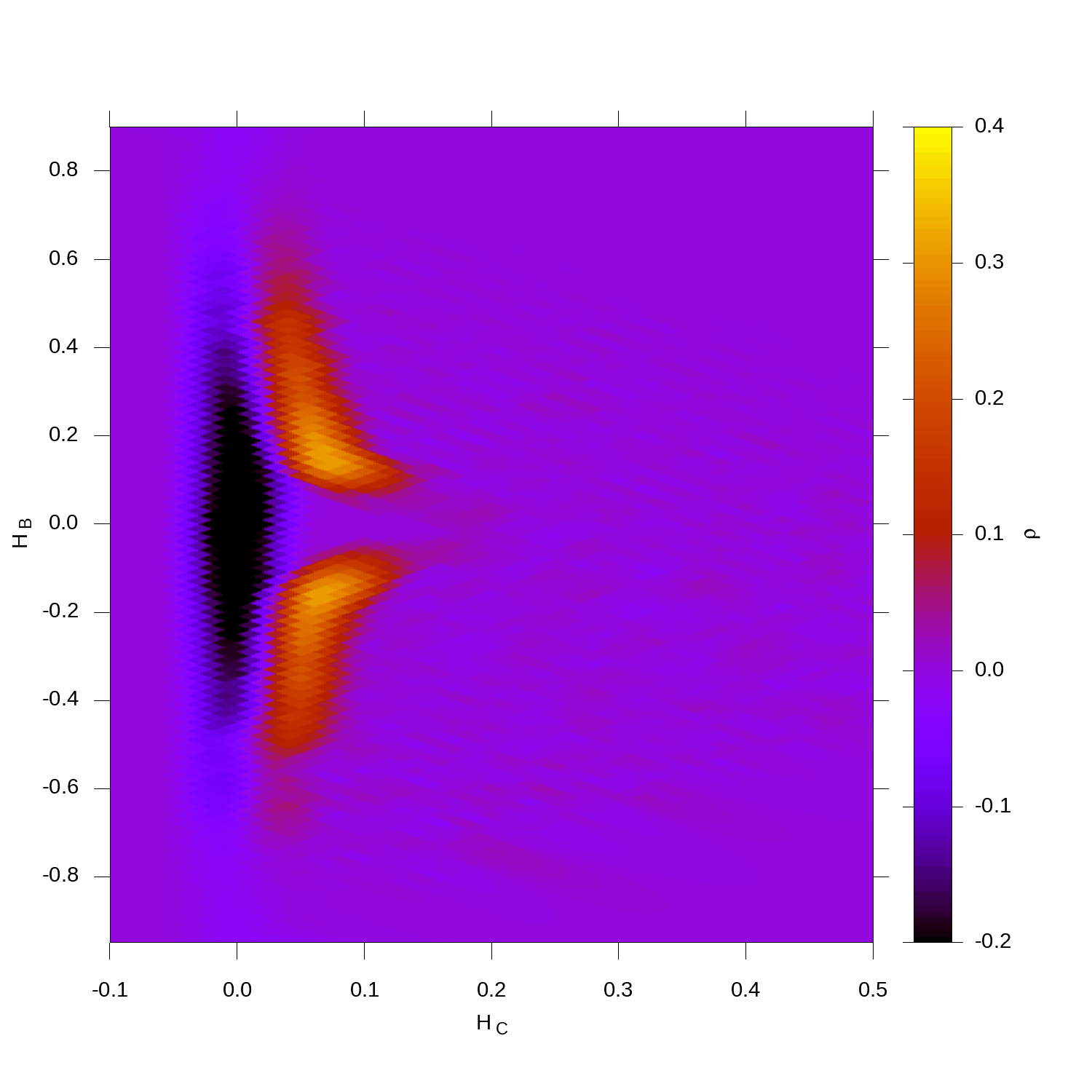}
\caption{FORC distribution for sample without array of defects, transformed coordinates.}
\label{dforc_sd1_0}
\end{figure}

The sample without defects presents perhaps the most characteristic signatures of FORC diagram structures for superconductors. Its FORC distribution (without applied current) is shown in figure \ref{dforc_sd1_0}. 

For plotting, it is more convenient to use a coordinate transformation, to the local coercivity and bias through the equations:

\begin{equation}
H _{C} = \frac{H - H _{r}}{2}
\label{Hc}
\end{equation}

\begin{equation}
H _{B} = \frac{H + H _{r}}{2}
\label{Hb}
\end{equation}

One can notice three isolated structures. A negative one, representing the reversible component, and two positive (approximately) symmetric, separated by a gap in $H _{B}$.

\begin{figure}[!htb]
\begin{center}
$\begin{array}{c@{\hspace{0.1cm}}c}
\multicolumn{1}{l}{\mbox{\bf (a)}} &
	\multicolumn{1}{l}{\mbox{\bf (b)}} \\
\includegraphics[width = 0.25 \textwidth]{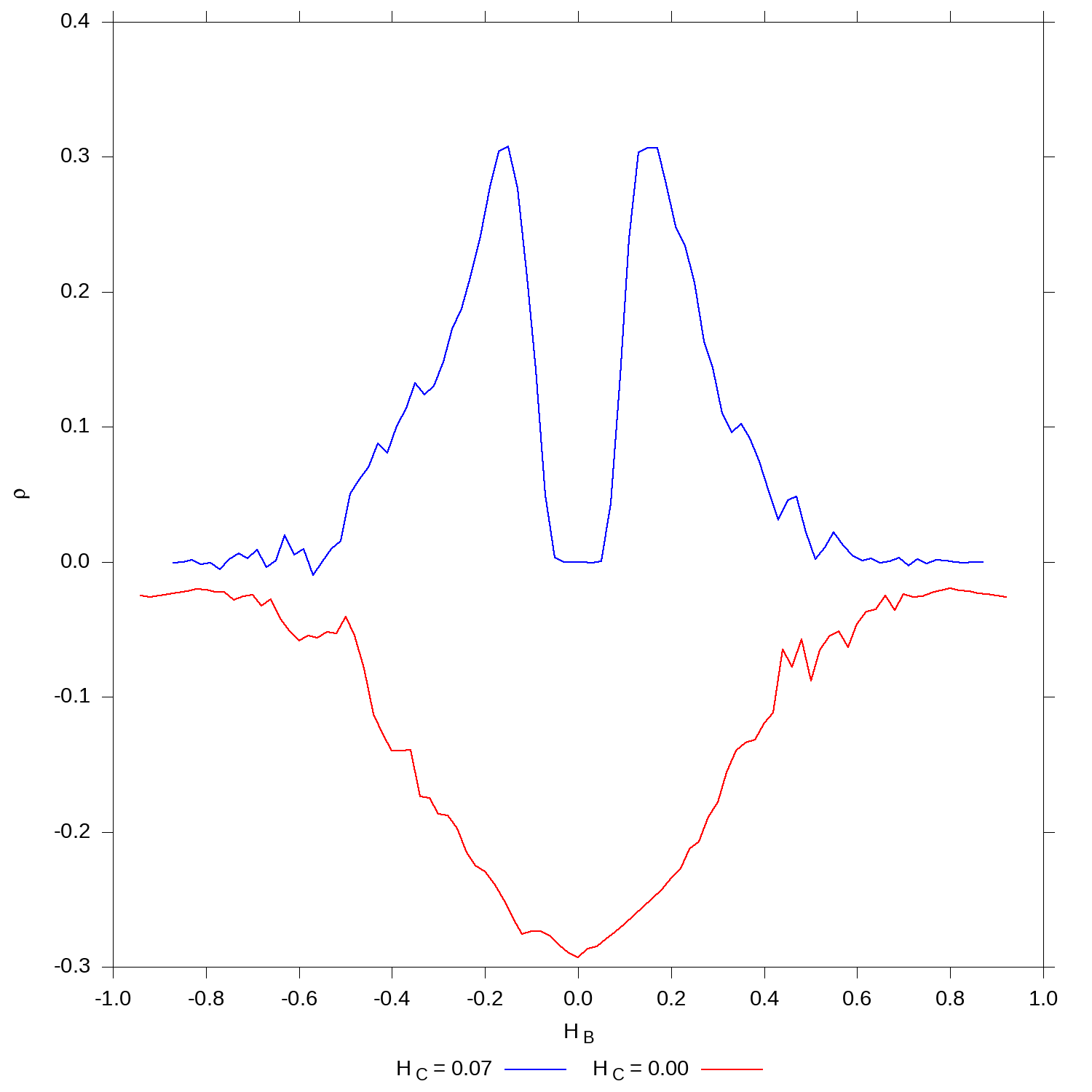} &
\includegraphics[width = 0.25 \textwidth]{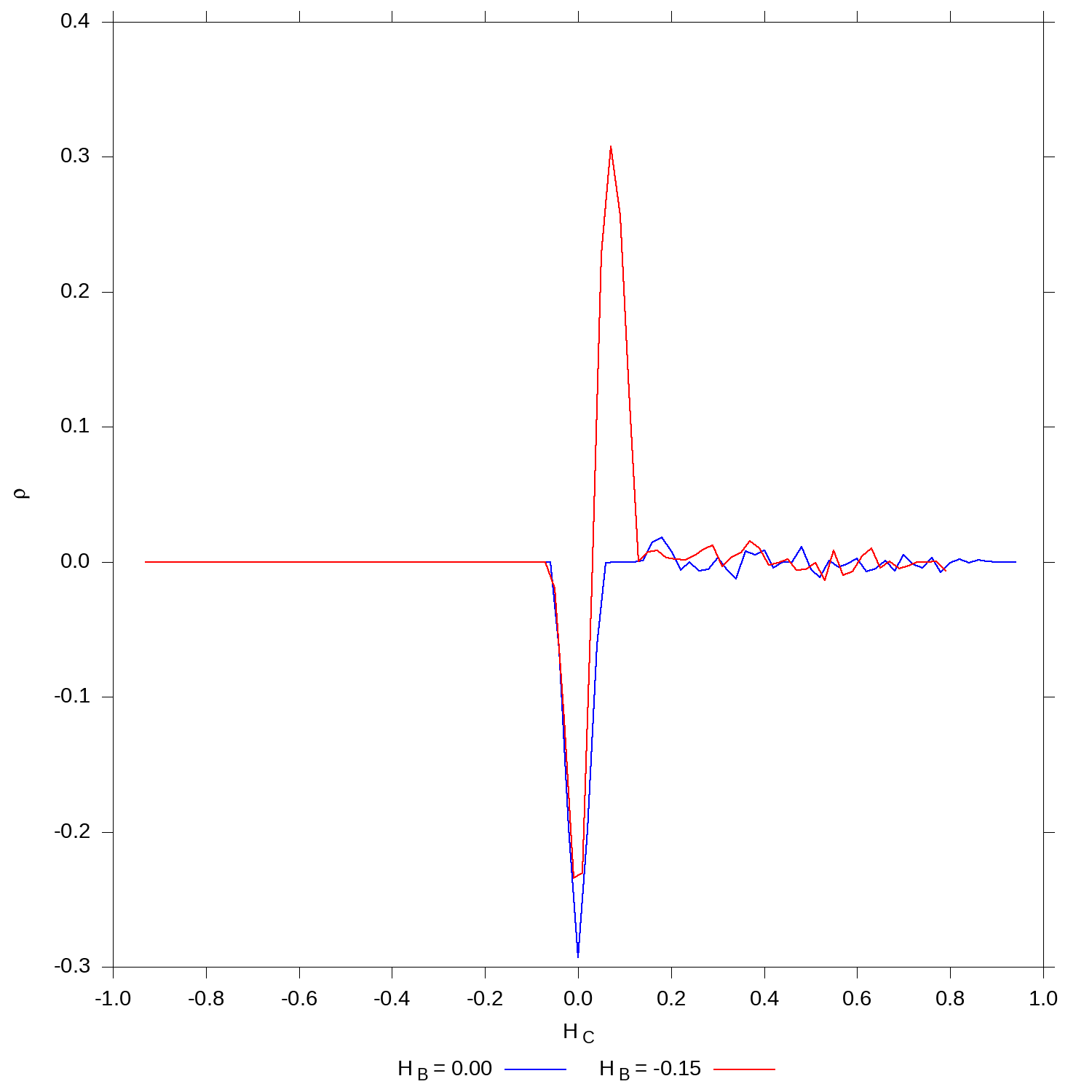} \\ 
\mbox{Constant $H _{C}$.} & \mbox{Constant $H _{B}$.}\\
\end{array}$
\end{center}
\caption{Cross sections with constant $H _{C}$ and constant $H _{B}$.}
\label{def_nets}
\end{figure}

The negative reversible component is a hallmark of superconductors, because of the Meissner effect. Both other structures are connected to vortex dynamics, since this simulated sample represents a type II superconductor. To see more details, one can look at the cross sectional profile curves, in figure \ref{def_nets}. 

One can notice that the reversible structure is (approximately) symmetric in $H _{B}$ and so are the positive peaks of the positive structures. The extension of the gap (from peak to peak) is $\Delta H _{B} = 0.32$. In graph {\bf b}, one can notice both the reversible valley (at both $H _{B} = 0$ and $H _{B} = -0.15$) and the peak at $H _{B} = 0.15$. Both structures are somewhat smoothed due to the regression fit mentioned before. 

One can notice from figure \ref{hysteresis_sd_0} that there is no horizontal asymptote to represent magnetic saturation. To use equations \ref{descending_branch}, \ref{ascending_branch}, a saturation value must be supplied. As it can be seen from \cite{Mayergoyz}, this value needs not to be connected to a horizontal assymptotes, but only to a reversible part of the hysteresis. This allows one to fix the $M _{S}$ value associated with a given field value. 

\begin{figure}[!htb]
\centering
\includegraphics[width=0.5\textwidth]{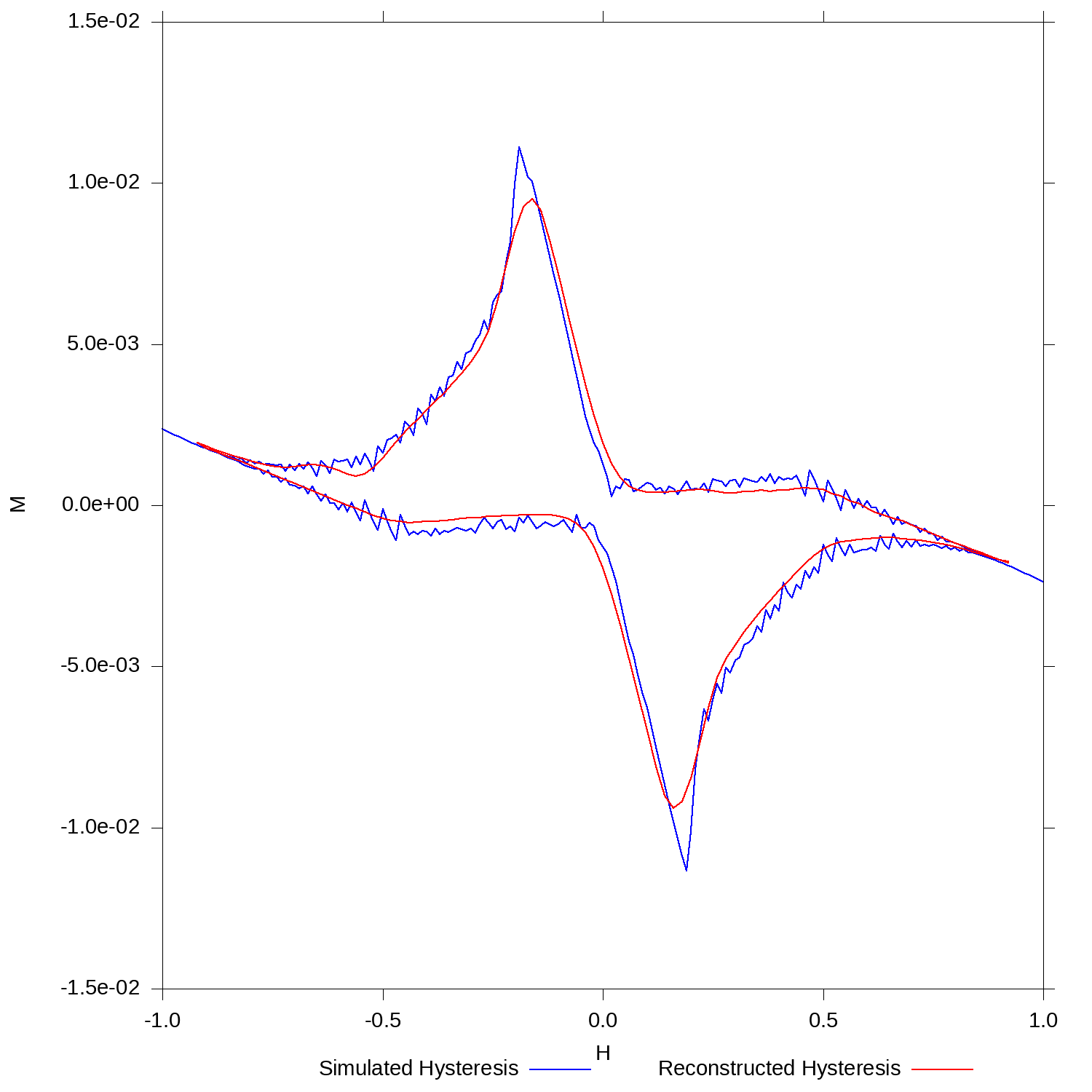}
\caption{Reconstructed hysteresis for sample without defects.}
\label{reconstructed_hyst_sd_0}
\end{figure}

Figure \ref{reconstructed_hyst_sd_0} shows this process applied to the first sample, using the FORC distribution data and the magnetic field values of $H _{app} = \pm 0.92$ to fix the magnetization saturation value.

Much of the residuals are due to the filtering in the FORC distribution calculation (eq. \ref{polynomial_forc}). Applying a paired t-test, one obtains $t = -1.8077$, with $\mbox{p-value} = 0.072$, meaning that the differences are not significant at the 5\% level, that is, the fit is acceptable at 5\% level. 

Reconstructing the hysteresis for both the positive and negative structures in figure \ref{dforc_sd1_0} is a little challenging, because the filtering has spread the structures in the x-axis direction and the magnetic field step (the resolution) does not allow a proper separation. One option to approximately separate both structures is to select a cut in $H _{C}$ (the x-axis in figure \ref{dforc_sd1_0}). 

\begin{figure}[!htb]
\centering
\includegraphics[width=0.5\textwidth]{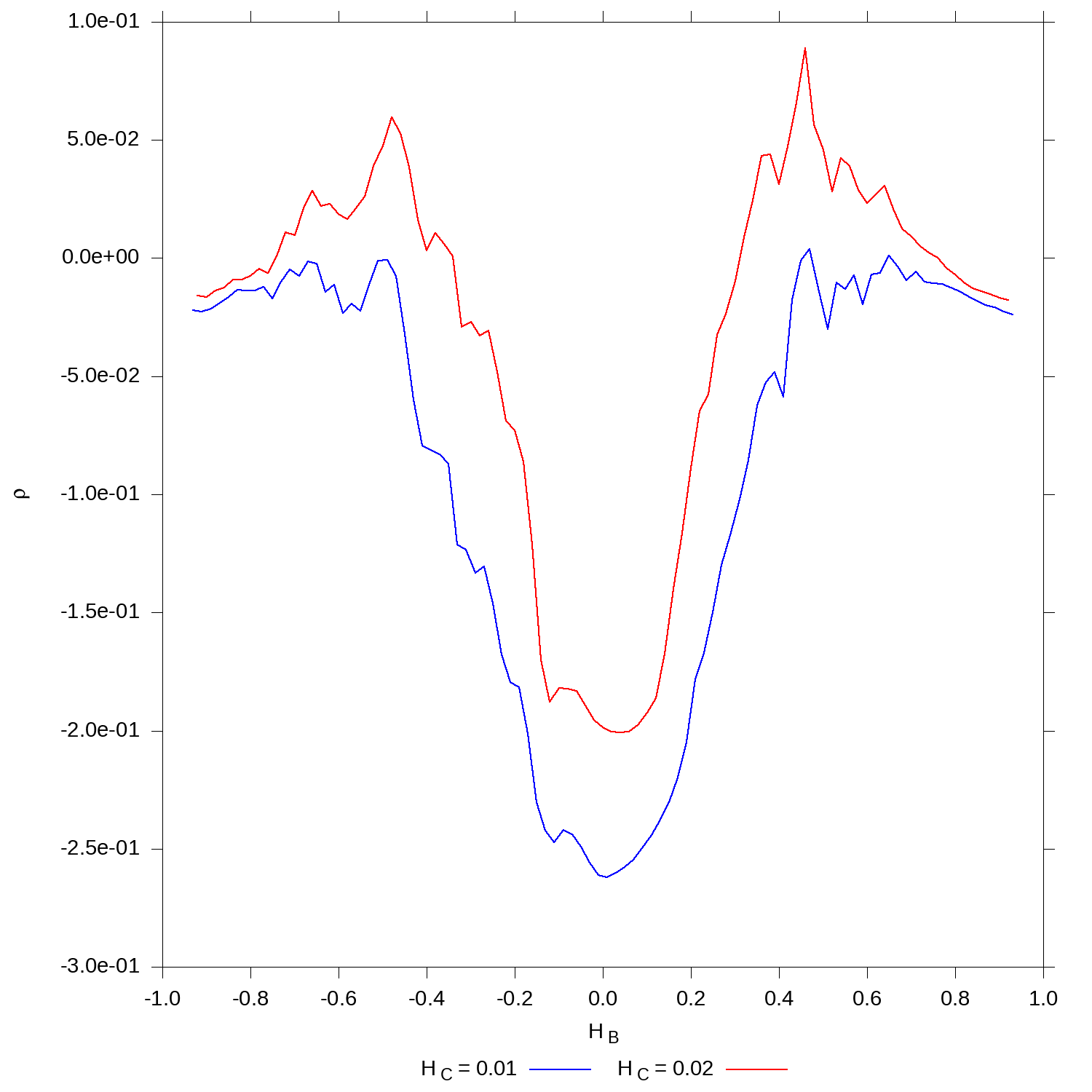}
\caption{Cuts in $H_{C}$ to separate positive and negative FORC structures.}
\label{prof_cut_1}
\end{figure}

Comparing the profiles for both possible cuts ($H _{C} = 0.01$ and $H _{C} = 0.02$, figure \ref{prof_cut_1}) allows one to infer a greater proportion of the positive structures contributing to the negative reconstructed hysteresis if $H_{C} = 0.02$ is chosen. Even $H _{C} = 0.01$ still has some residual positive structure. So the cut will be chosen as $H _{C} = 0.01$, that is $H _{C} \leq 0.01$ will be attributed to the negative structure, whereas $H _{C} \geq 0.02$ to the positive structures (recalling that 0.01 is the resulting resolution in $H _{C}$). This also means some contribution of the negative structure to the positive ones.

Using those cuts, the positive and negative stuctures lead to the following reconstructed hysteresis curves.

\begin{figure}[!htb]
\centering
\includegraphics[width=0.5\textwidth]{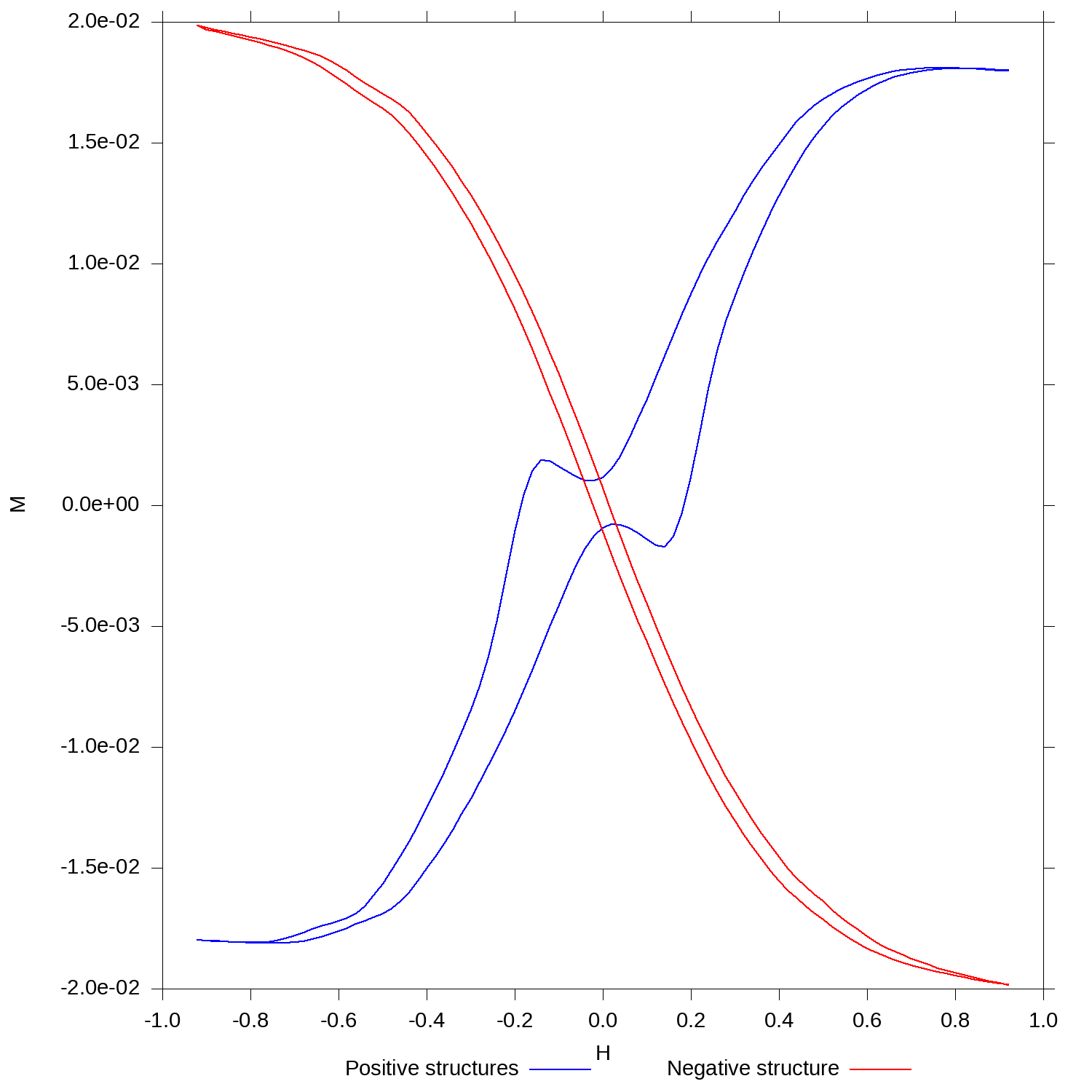}
\caption{Reconstructed hysteresis curves from the positive and negative structures in the FORC diagram.}
\label{comp_sd1_0}
\end{figure}

One can notice that the negative structure is (only approximately) reversible. This may be attributed partially to the filtering process and corresponding numerical smoothing. Another possible contributing factor is the accuracy in TDGL methodology representing normal metals, which is the case when applied to fields above the critical value (in this case with normalized units and at the temperature studied, $H _{c2} = 0.5$). But it is also consistent with a reversible phase that is inherently not separable from the hysteretic, perhaps due to the penetration of the magnetic field in the sample with penetration length $\lambda$. Connected to this, the quality of the reversibility approximation should increase as the sample size increases, up to when penetration effects become negligible. To be completely sure, one would have to increase the resolution of the magnetic field and use a bigger sample in an experimental or numerical setting.

So, for the remaining analysis, the negative valley at $H _{C} = 0$ will be referred to as reversible component (even if it actually is only an approximate reversibility or the structure is not completely separable). 

With that in mind, one can start analysing two hallmarks of the superconducting hysteresis (type II), particularly connected to FORC analysis. The negative reversible component (behaving like a diamagnetic one) and the more usual hysteretic component, with basically two plateaux from $0$ to $-0.2$ in the descending branch, and from $0$ to $0.2$ in the ascending one. Both are followed by a more steep curve than before, up to saturation.

This is connected to vortex dynamics. Taking the ascending branch, for example. From negative saturation up to 0 the negative vortices, that had penetrated the sample going to negative saturation, are being eliminated. The plateau corresponds to the sample still being in the Meissner state, as shown by direct vortex mapping in \cite{kato}. After that, the positive vortices start entering and moving inside the sample. This permanence in the Meissner state is responsible for the gap in the positive symmetric structures in the FORC diagram.

Since one of the hallmarks of the superconducting (type II) FORC distribution is connected to vortex dynamics, it is important to uderstand how this is expected to change when the vortex dynamics is changed. For that, it is worth studying the characteristics of hybrid samples, where vortex pinning structures of different material are introduced in the sample. This also has the potential to show what is to be expected from more complex hybrid structures such as supeconducting spin valves, multilayer, etc. 

\begin{figure}[!htb]
\begin{center}
$\begin{array}{c@{\hspace{0.1cm}}c}
\multicolumn{1}{l}{\mbox{\bf (a)}} &
	\multicolumn{1}{l}{\mbox{\bf (b)}} \\
\includegraphics[width = 0.25 \textwidth]{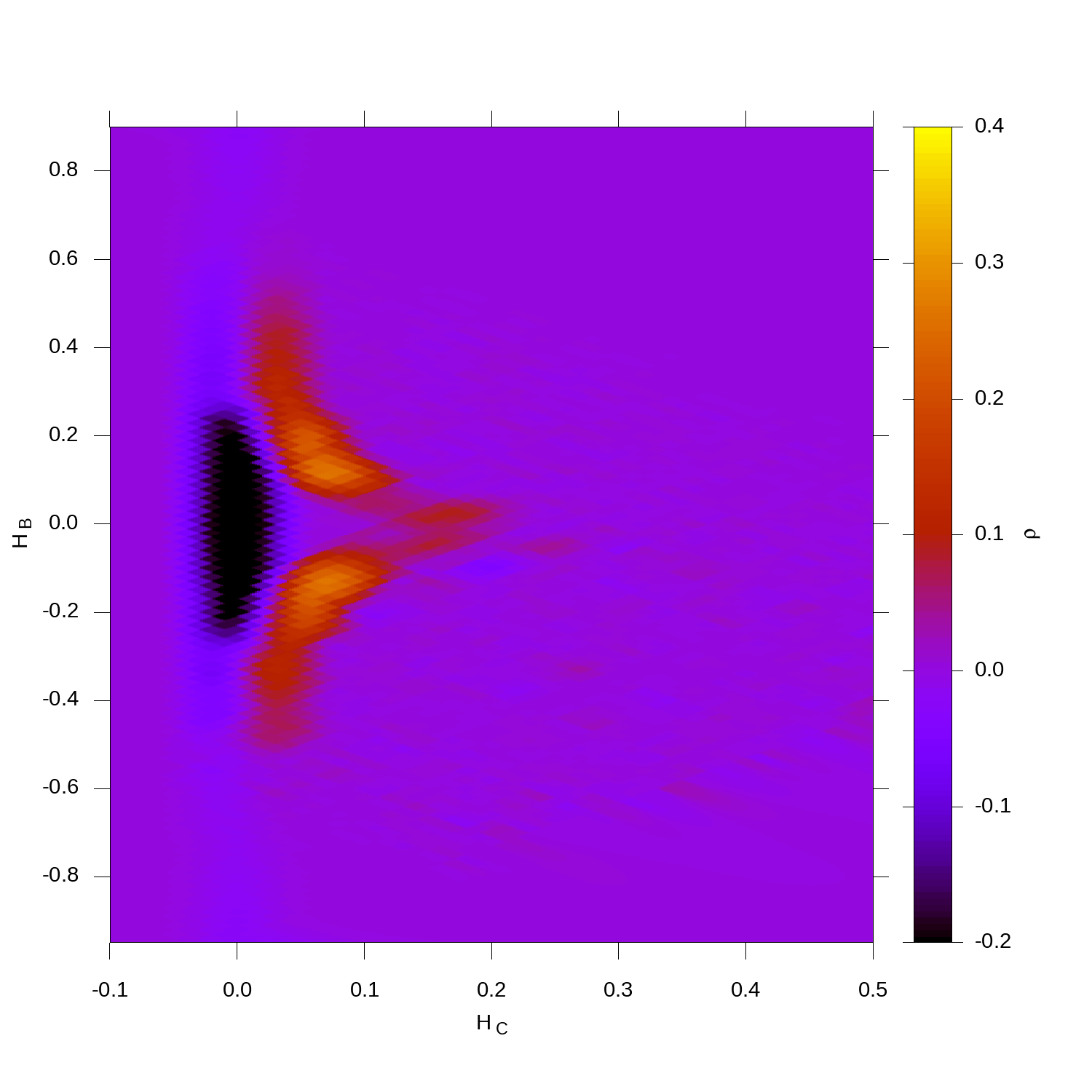} &
\includegraphics[width = 0.25 \textwidth]{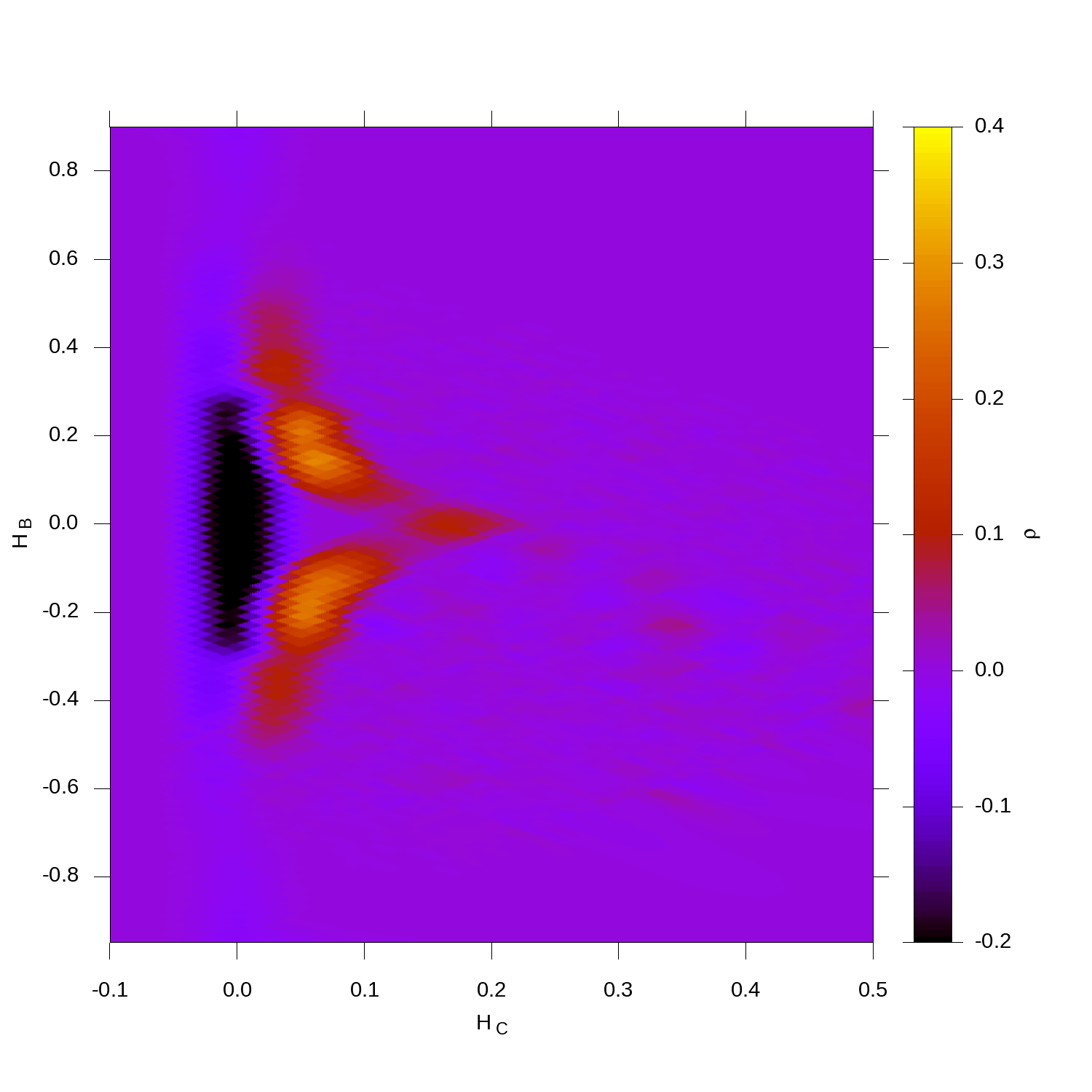}	 \\ 
\mbox{Square} & \mbox{Triangular} \\ [0.2cm]
\multicolumn{1}{l}{\mbox{\bf (c)}} &
	\multicolumn{1}{l}{\mbox{\bf (d)}} \\
\includegraphics[width = 0.25 \textwidth]{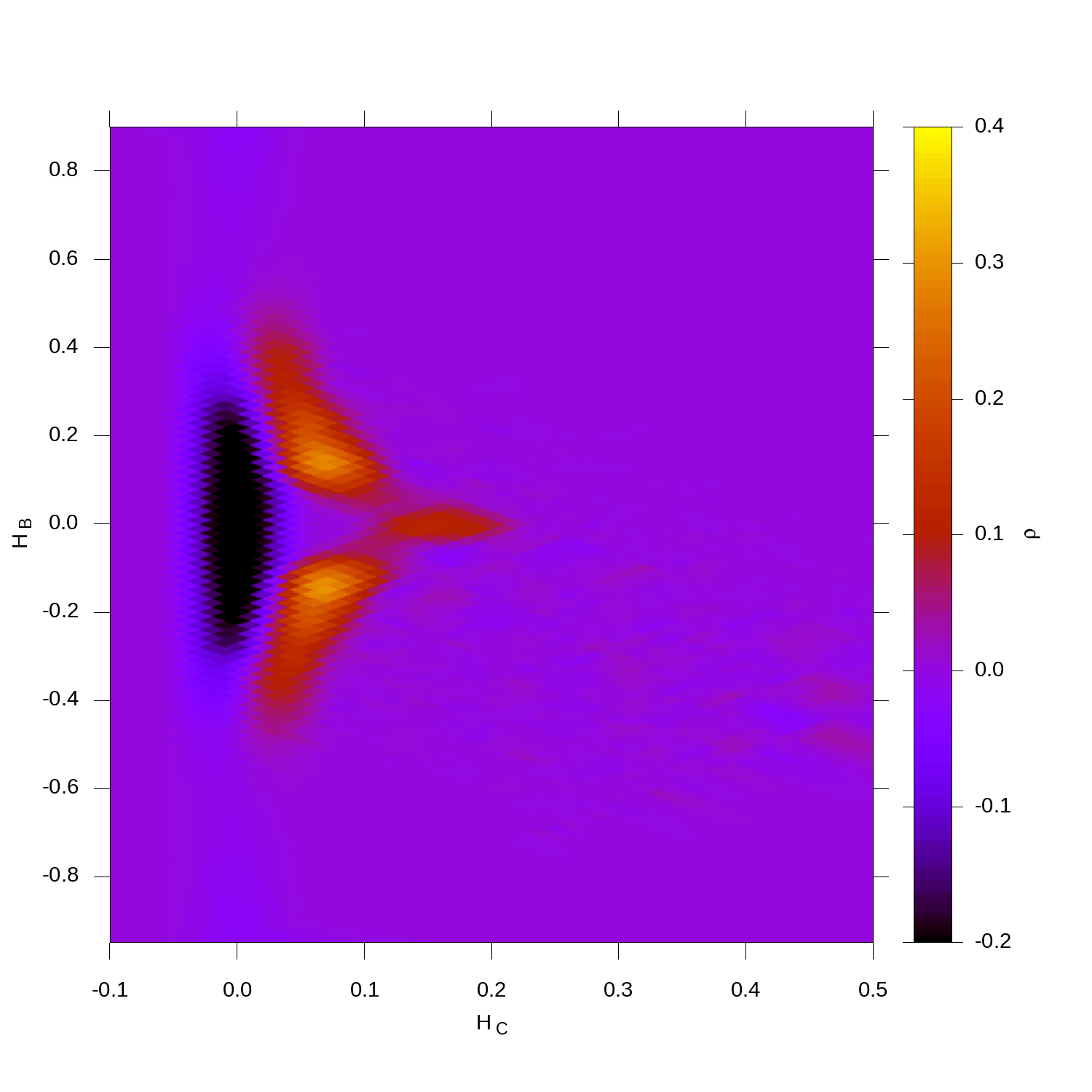} &
\includegraphics[width = 0.25 \textwidth]{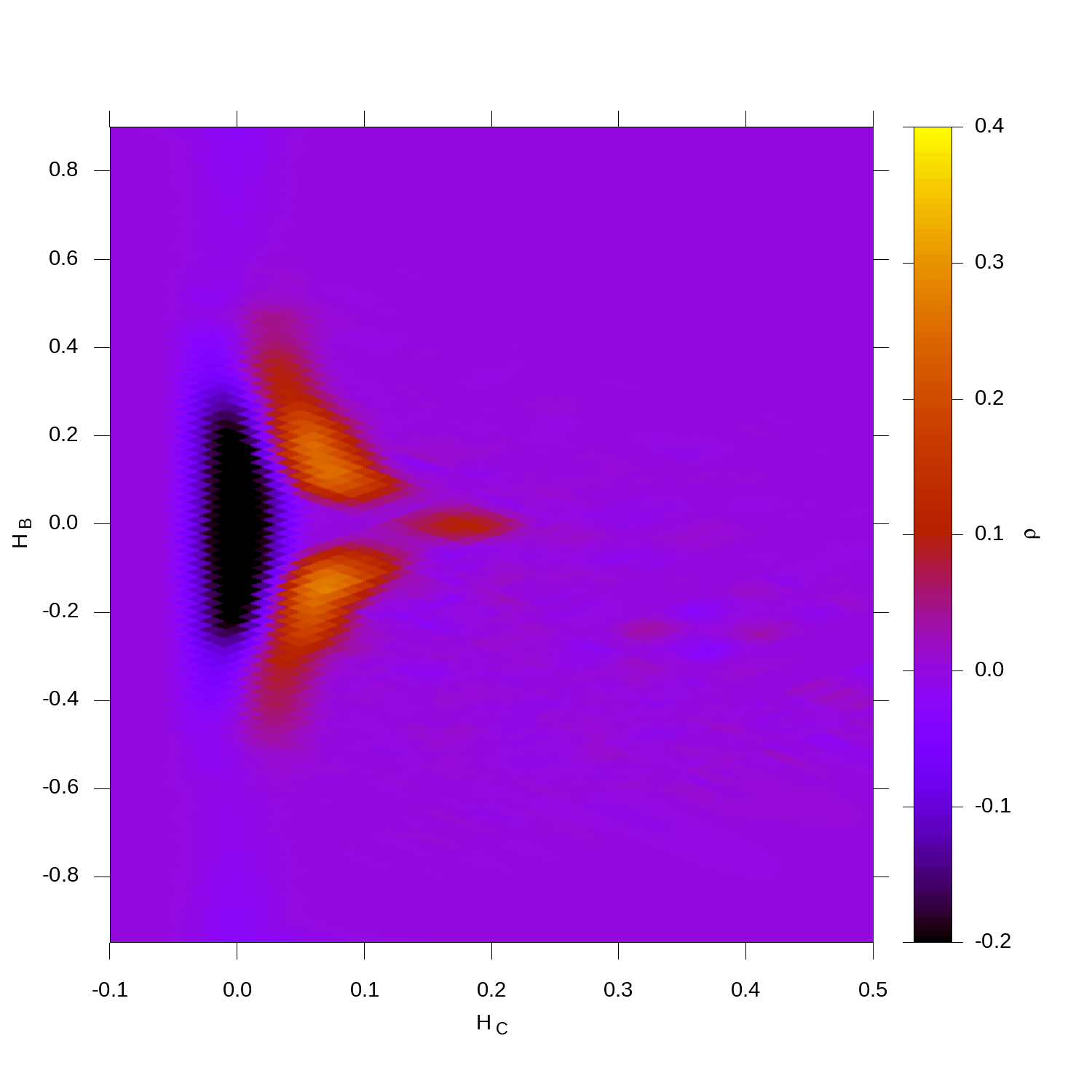}	 \\ 
\mbox{Center concentrated} & \mbox{Border concentrated}
\end{array}$
\end{center}
\caption{FORC diagrams for samples with pinning arrays.}
\label{forcs_i0}
\end{figure}

Using a set of pinning arrays that were shown to increase critical current due to vortex pinning (\cite{lascio1}), one obtains the set of FORC diagrams in figure \ref{forcs_i0}.

The first noticeable feature is a new positive structure, for all samples. This structure is smaller than the two other positive ones, and is concentrated around $H _{B} = 0$, between $H _{C} = 0.15$ and $H _{C} = 0.2$. It tends to close the gap between the two other positive structures, and it seems like it almost does that for both the triangular and the center concentrated arrays. For the square array, this structure is less defined, and for the border concentrated it seems more clearly isolated from the others.

The pinning arrays modify the vortex dynamics. So this structure is related to vortices remaining even when the sample is in Meissner state. To better see this, one can observe the reconstructed hysteresis for the three sets of structures: the negative, the two positive closer to the reversible negative one (analogous to those in the sample without pinning centers) and the new positive one. For that, the most convenient is the border concentrated sample, where the structure is more clearly isolated, this third structure will be referred to as island. 

Another interesting detail is that both the square and the triangular arrays show more evidently some segmentation of the positive structures, like multiple peaks of varying intensity, perhaps due to matching fields.

Is is important to mention that once again, even for the border concentrated sample, the structures are not completely separable using a cut in $H _{C}$, given the field resolution applied in the simulations. This may be different with more resolution, perhaps in a future experiment. Even so, in an attempt to minimize the influence of the other structures over the island, values smaller than or equal to $H _{C} = 0.13$ (but still greater than the lower limit $H _{C} = 0.01$) will be attributed to the positive structures and greater than 0.13 to the island.

\begin{figure}[!htb]
\centering
\includegraphics[width=0.5\textwidth]{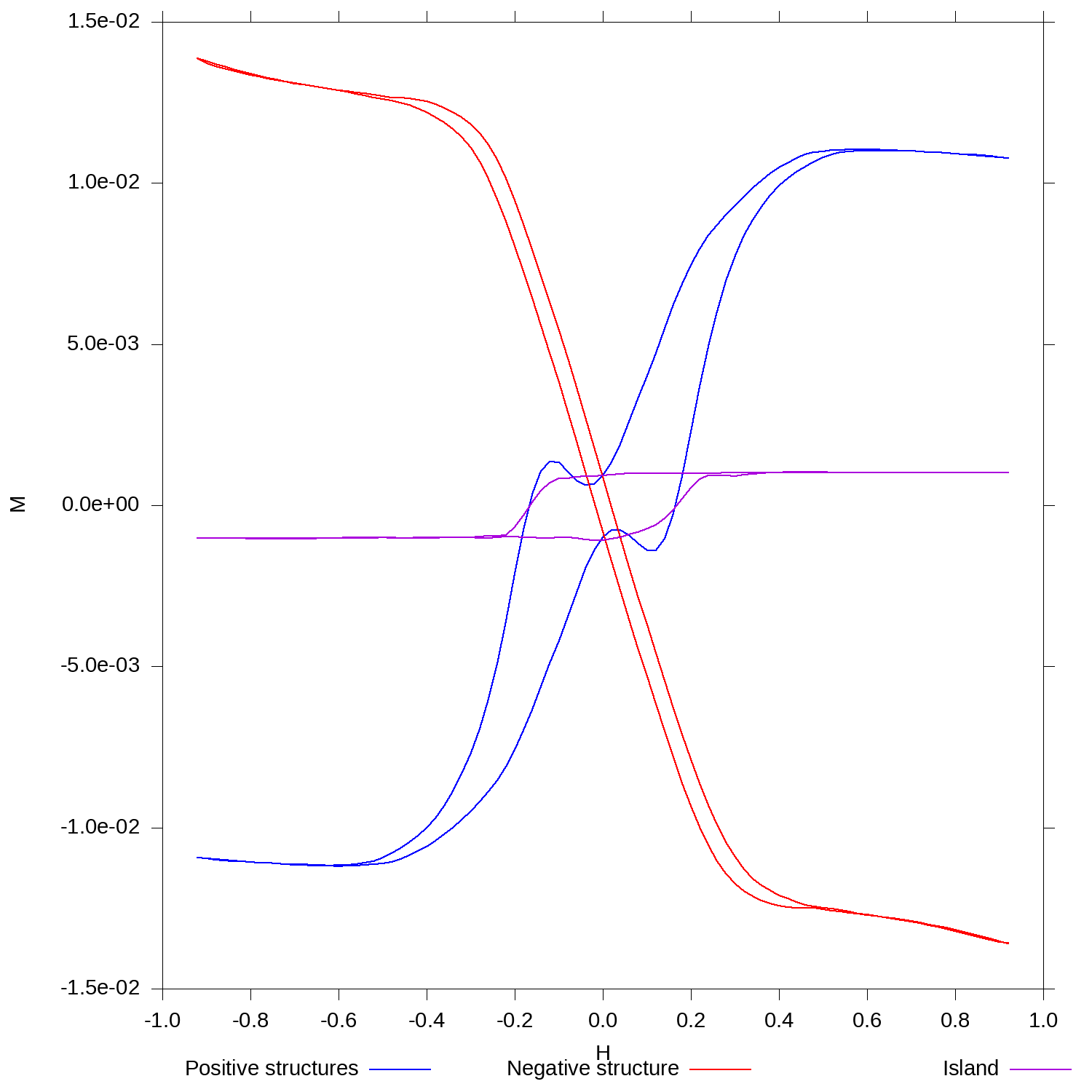}
\caption{Separated reconstructed hysteresis curves from the border concentrated sample.}
\label{comp_bc1_0}
\end{figure}

Using those cuts, the island, positive and negative stuctures lead to the separated reconstructed hysteresis curves of figure \ref{comp_bc1_0}.

After saturation, when the field starts reversing, part of the vortices is kept pinned to the array. From the figure, one can notice that the hysteresis from the island reverses the magnetization when the hysteresis from positive structures starts moving towards saturation, that is, when (positive or negative) vortices begin entering the sample again. These new vortices annihilate the vortices that had been kept due to the pinning array (the island), which quickly starts accumulating oposite vortices, now aligned with the orientation of the positive structures.

In a way, the hysteresis from the island is very similar to a ferromagnetic one. That can have two possible uses. One is that the pinning array presents a FORC signature similar to a ferromagnetic one, which means that it can be a confounding factor when identifying phases in a hybrid sample through FORC analysis. The other is that since the pinning of vortices is connected to increased critical currents, some mechanism that increases this island, therefore keeping more vortices pinned when the hysteresis is reversing, could result in improved critical currents. Maybe an ferromagnetic or hybrid (alternating ferromagnetic and normal metal sites) could result in increased island hysteresis amplitude and hence higher critical currents, if the ferromagnetic sites are introduced without destoying the superconductivity and the proximity effect is not too relevant to change the FORC structure (in that case a new analysis is necessary, but the critical current could still be improved). 

To get a little more insight into this matter, one can try to correlate critical currents to the intensity and dimensions of the island and the magnetometric parameters of its hysteresis.

\begin{table}[h]
\caption{Magnetometric characteristiscs of the island for samples that present it in FORC diagram.}
\vspace{0.2cm}
\label{island_no_I} 
\centering 
\begin{tabular}{cc}
{\bf {\it Sample}} & $\mathbf{\rho _{max}}$\\ \hline
{\bf{Square}} & $0.897 \cdot 10 ^{-1}$\\ \hline
{\bf{Triangular}} & $1.02 \cdot 10 ^{-1} $\\ \hline
{\bf{Border Conc.}} & $1.05 \cdot 10 ^{-1} $\\ \hline
{\bf{Center Conc.}} & $1.14 \cdot 10 ^{-1} $\\ \hline
\end{tabular}
\end{table}

The maximum value of $\rho$ for the island increases as the pinning of vortices increases, and a few metrics related to critical current improvement \cite{lascio1} show that the higher the peak of the FORC distribution for the island, the greater the critical current improvement is, with respect the pure superconductor.

The bottom line is that a critical current improvement for the sample can be inferred from purely magnetometric measurements.

To explore the magnetometric parameters of the reconstructed hysteresis for the island would demand more field resolution, since even for the most isolated island (that for the border concentrated sample) there is still some mixture between the island and the two other positive structures. Parameters like remanence of the reconstructed hysteresis could also be correlated with critical current improvement metrics, but that would demand minimizing the residual contribution of the other positive structures to the hysteresis of the island, as well as increased magnetic field resolution, to allow for a better separation.

But to understand a little better how the samples behave with an applied current, as expressed by the FORC distribution, more simulations were performed, including a value of current that reached critical values for some values of applied field for all samples, $I _{app} = 0.4$.  

\begin{figure}[!htb]
\centering
\includegraphics[width=0.5\textwidth]{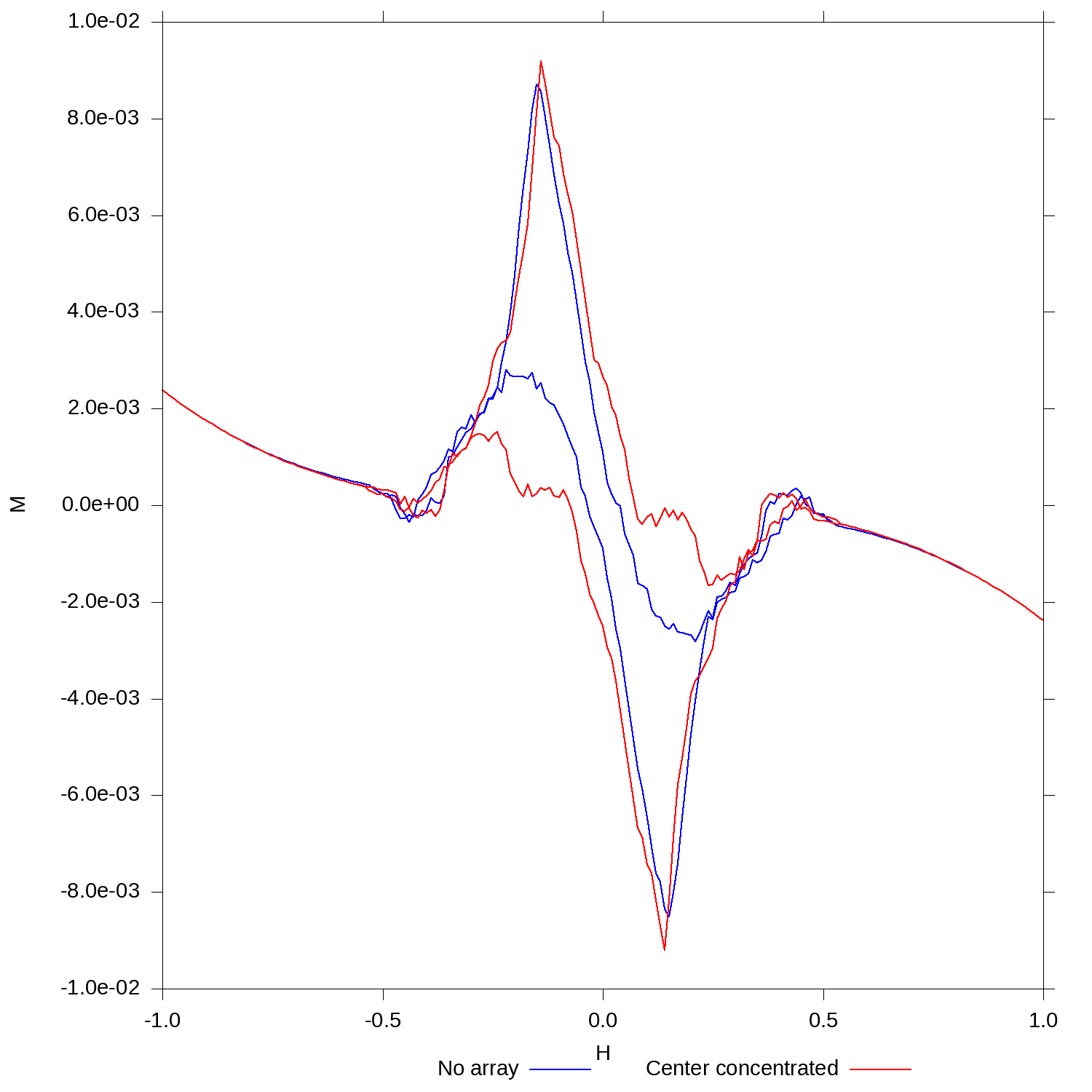}
\caption{Hysteresis for samples without pinning array and with a center concentrated array with applied current.}
\label{hyst_cur}
\end{figure}

The first changes to be noticed are evident from the hysteresis curves. Figure \ref{hyst_cur} shows two examples, for the pure superconductor and for the center concentrated sample. The amplitudes (represented by the remanence, for instance) are smaller and the (approximately) reversible phase is reached for smaller applied field values. Furthermore, this reversible behaviour now has the peculiar characteristic of reversing the sign as the applied field is moved further towards saturation. This is not actually reversible (at least for the entirety of the paramagnetic-like behavior), since besides being more noisy (which alone wouldn't be enough to destroy revesibility), it presents one mini loop at each extremity. 

Also relevant is the fact that the sample without pinning array gets thinner around $H = 0$, since it has no means of retaining the vortices while entering the Meissner behaviour region.

\begin{figure}[!htb]
\begin{center}
$\begin{array}{c@{\hspace{0.1cm}}c}
\multicolumn{1}{l}{\mbox{\bf (a)}} &
	\multicolumn{1}{l}{\mbox{\bf (b)}} \\
\includegraphics[width = 0.25 \textwidth]{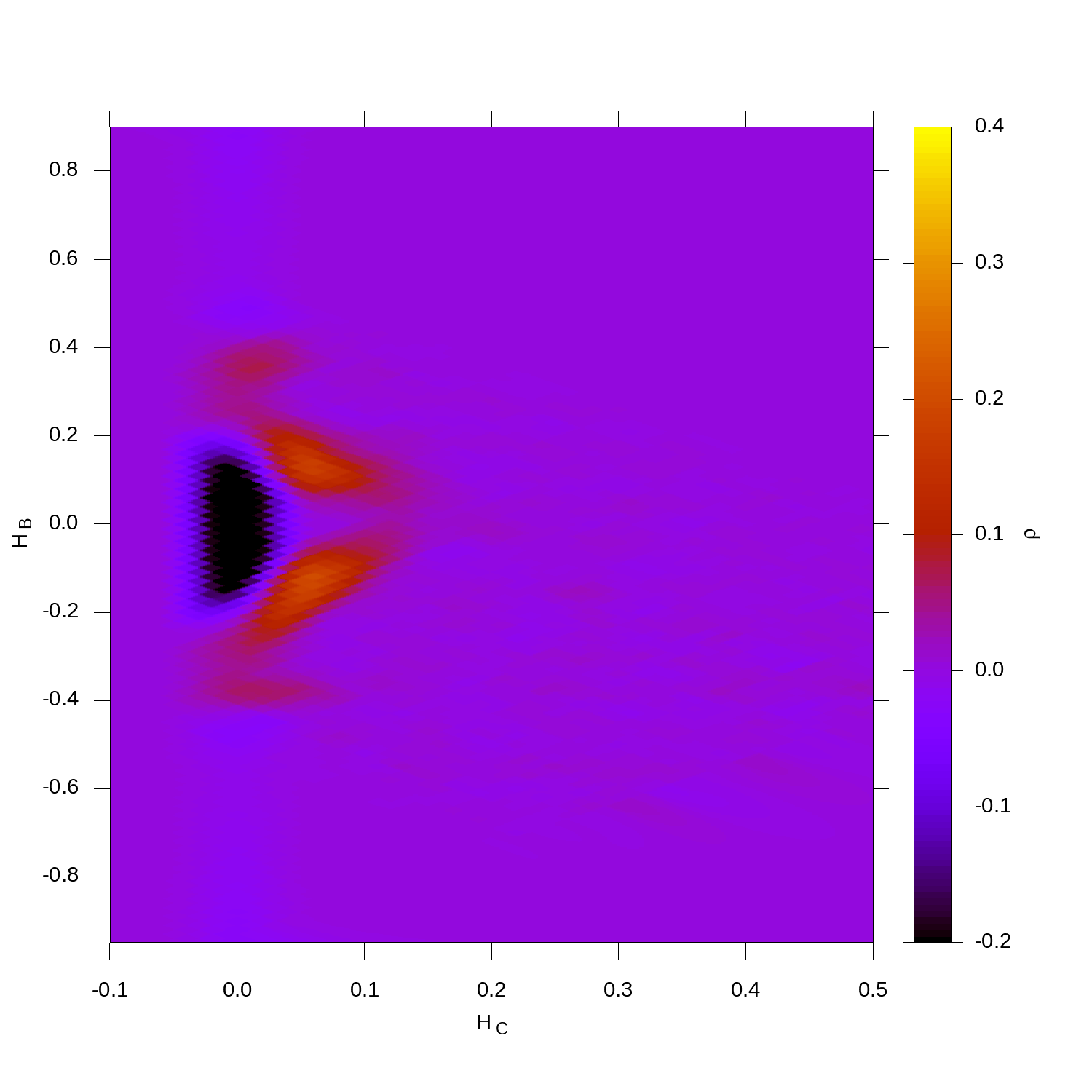} &
\includegraphics[width = 0.25 \textwidth]{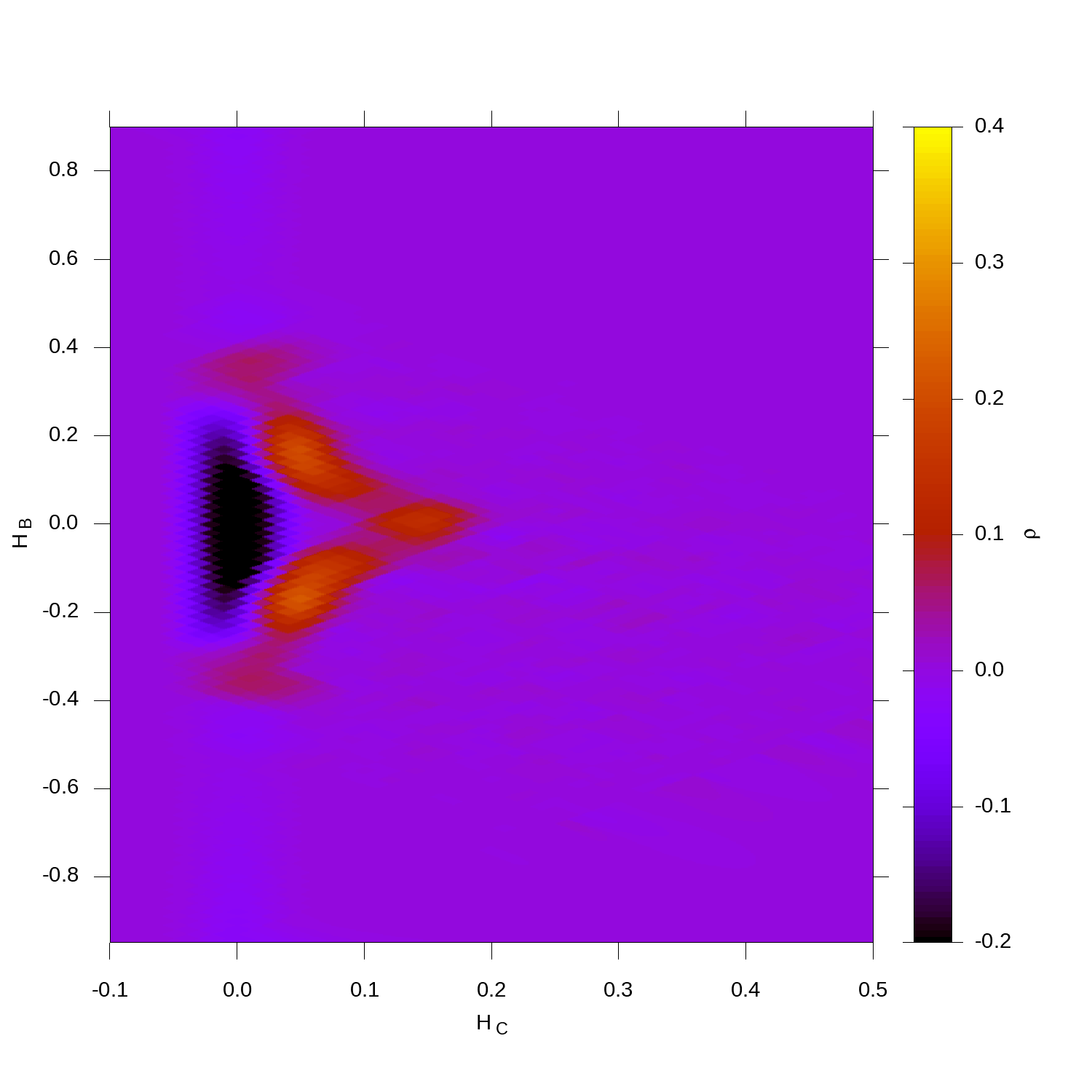} \\ 
\mbox{No pinning array.} & \mbox{Centre concentrated.}\\
\end{array}$
\end{center}
\caption{FORC diagrams for samples with applied current.}
\label{forc_cur}
\end{figure}

This is reflected in the FORC diagrams as well, shown for both samples in figure \ref{forc_cur}. 

The island, associated with this retaining of vortices is kept qualitatively intact in the center concentrated sample. The intensity of the FORC distribution is smaller for both samples.

But the most interesting qualitative change is the appearance of two new structures at the edge of the negative region, for both samples. This is even more interesting because they are positive and effectivelly make the (approximate) reversible ridge to reverse its sign for some intervals in $H _{B}$. That is connected to the paramagnetic-like behavior, in sequence with the diamagnetic one when evolving the applied field, which is apparent from the hysteresis itself. 

That the sample can behave like a paramagnet (except for the intriguing two mini-loops) and as the field evolves further it reverses to the more ususal diamagnetic behaviour is certainly connected to the vortex dynamics. The most likely way is that this region represents an example of the flux flow regime under magnetometric measurements. Another possibility is that this represents an accelerated region of vortex creation and annihilation. The analysis of the detailed vortex dynamics mechanism in this region demands more studies, but the most important point is that its behavior is captured by the FORC diagram as these new positive structures overlapping with the negative structure.

\section{Conclusion}

FORC analysis is a technique that continuously gain applications as more people apply it to various hysteretic systems and its interpretation evolves. Currently, despite its origin being connected to Preisach modelling, it has evolved to a magnetometric technique that enables the researcher to identify fingerprints of magnetic phases, particularly important in complex hybrid systems. It also leads to insights about interactions between phases, as can be seen from the capability of separating hysteresis curves for different phases.

This type of insights about the interactions and identification of phases can be invaluable for new more complex materials. To employ this technique with all the advantages, among which is that it is very simple to implement, to new materials would have three great outcomes. First, the identification of the structures present in the new materials would help to identify how much these contribute to the magnetic behavior of hybrid samples, along with insights about the interactions within the sample. Second, this identification and correlation with interaction mechanism (like vortex dynamics) could help identify mechanisms of interaction in some materials, like spin-pumps and spin currents, ruling out some potential explanations and keeping others. Third, it would allow better sample design, considering desired characteristics and the outcome of purely magnetometric measurements. As a bonus, with more specific materials being investigated with FORC analysis, more people would be aware of it, testing, contributing and improving the interpretation of the technique.

This paper expands the application of the FORC analysis to superconductors and hybrid materials (S/N), where a wealth of new applications can be made, such as thin films, multilayers, spin valves. It starts adapting the definition of saturation to make it possible to be used with superconductors. Employing the definition of FORC distribution, the paper identifies characteristic features of superconductors, like the negative (approximately) reversible ridge, positive structures connected to vortex dynamics, a positive island related to vortex pinning (correlated also with increased critical currents) and new structures (mini-loops and positive intervals of the (approximate) reversible ridge) with applied currents. These contributions will make it possible for new experimental work using FORC analysis to advance in the interpretation of the interactions among the components of the sample, being clear now what is expected of a superconductor with the FORC distribution.

\section*{Appendix}

Despite in common use, the fitting of a polynomial to obtain the distribution is not necessarily the best choice for the numerical calculation of the distribution. 

With discrete differences, one obtains a more noisy distribution, but a clearer picture of what is happening in the sample. 

To see the differences, a comparison of the distributions calculated differently are in order.

\begin{figure}[!htb]
\centering
\includegraphics[width=0.5\textwidth]{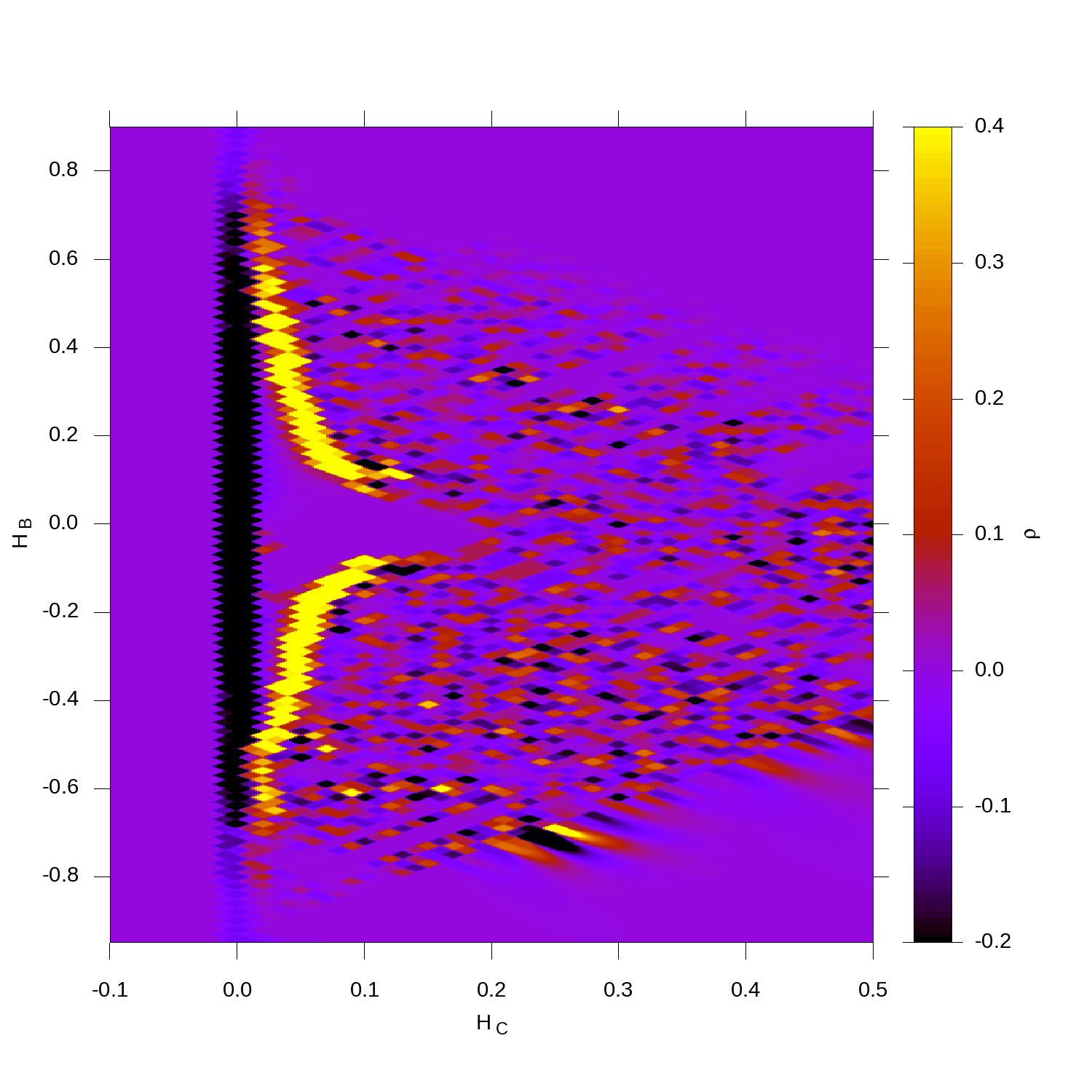}
\caption{FORC distribution calculated with discrete differences for the sample without pinning array.}
\label{comp_dist_007}
\end{figure}

One can see, comparing figure \ref{comp_dist_007} with figure \ref{dforc_sd1_0}, structures that are less spread but more intense, without the filtering of the polynomial. It becomes clearer that the negative and positive structures are not separable. It gives more noise, that is, the variance is bigger, but not biased values. 

\begin{figure}[!htb]
\centering
\includegraphics[width=0.5\textwidth]{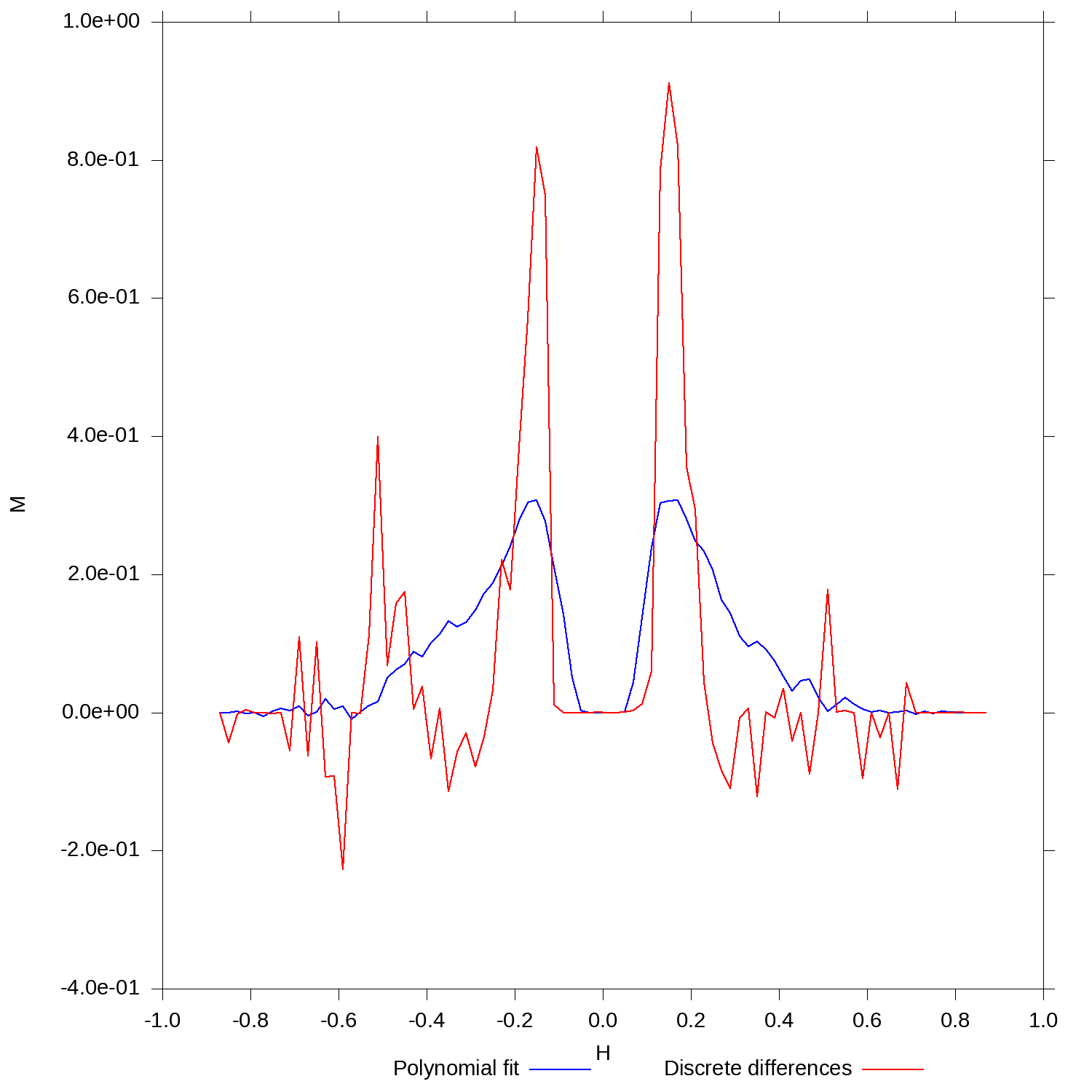}
\caption{Profile curves for $H _{C} = 0.07$, comparing both methodologies.}
\label{comp_pro_007}
\end{figure}

That is, the best estimate for the peak in positive structures would be 0.9, not 0.3 as seen from the filtered distribution.

The relevance of the noise would be in hypothesis testing, for instance, to know how significantly different from zero a particular feature is. Otherwise, it gives more precise results, even for reconstructed hysteresis.

The conclusions of this article were checked with this other methodology as well, but the results were presented in the more familiar form, with the polynomial adjustment. It should be borne in mind, however, that this methodology gives a clearer picture and should be consulted in case of doubt or to check the conclusions, at least qualitatively.


\end{document}